# On the importance of prismatic/basal interfaces in the growth of $(\bar{1}012)$ twins in hexagonal close-packed crystals

B. Xu[a,b], L. Capolungo[c], D. Rodney[a]

[a] Laboratoire de Science et Ingénierie des Matériaux et Procédés, Grenoble Institute of Technology, CNRS, UJF, 38402 Saint Martin d'Hères, France.

[b] Laboratory of Advanced Materials, School of Material Science and Engineering, Tsinghua University, Beijing 100084, PR China.

[c] George Woodruff School of Mechanical Engineering, Georgia Institute of Technology Lorraine, UMI 2958 Georgia Tech–CNRS, 57070 Metz, France

The growth process of $(\bar{1}012)$ twins is studied in Magnesium using atomistic simulations. It is shown that a specific interface, which places face-to-face prismatic and basal planes, plays an important role. This interface has a low energy forming a cusp in the orientation-dependent interface energy of a twinned bicrystal. This interface appears in several published twin structures and for instance accommodates the large deviations of twin interfaces from $(\bar{1}012)$ planes reported recently [Zhang *et al.*, Scr. Mater. 67 (2012) 862].

Twinning, a mechanism of plastic deformation by which a crystal sub-domain is reoriented by a mirror symmetry operation with respect to a twinning plane, is of prime importance in low-symmetry materials and in particular in hexagonal-close-packed (hcp) metals [1]. The mechanism of twinning is typically seen as the sequence of three phenomena: (1) nucleation, (2) transverse propagation and (3) thickening (i.e. growth of twins perpendicularly to the twin interface). As thoroughly discussed in the seminal work of Partridge [2], twinning in hcp metals induces a shuffling of atoms, implying a more complex structure of the twinning dislocations as for instance in face-centered cubic crystals where no shuffling is needed. A consequence is that hcp twinning dislocations extend over several atomic layers [3], and in such a case, are referred to as disconnections [4]. Twinning dislocations form steps on the twinning plane thus increasing or decreasing the twin thickness as they glide along the twin interface, much like kinks on high-Peierls stress dislocations [5].

Since the early 1950s [6], twinning has been the subject of several important studies aiming at understanding its activation and influence on other aspects of plasticity (for instance latent effects and transmutation). Focusing here on the onset of nucleation and propagation, early analytical models have proposed different scenarii for the heterogeneous nucleation of twin domains in hcp systems, including the pole mechanism of Thompson and Millard [7] and the more general slip dislocation dissociation mechanism of Mendelson [3]. In recent years, atomistic simulations have complemented these studies by revealing several potential dislocation-based structures of twin nuclei. Most of the works considered the $(\bar{1}012)$ twin interface that is commonly activated in hcp metals. In Ref. [8], a twin nucleus was proposed, composed of a dipole of partial dislocations, which spontaneously forms a twin embryo in-between the two partials. However, the growth process of this embryo was not studied in details.

With regards to twin growth (i.e. transverse propagation and thickening), the thickening mechanism, i.e. the motion of the twin perpendicular to the twinning plane, has attraction a lot of attraction. It has been shown [9,10] that thickening is mediated by the glide of twinning dislocations along the twin planes, although there is some debate about an alternative mechanism involving atomic shuffling exclusively [11-13]. Interestingly, the process of transverse propagation of the twin, i.e. its extension in the twinning direction, has received far less attention and is the subject of the present study.

Twin growth is studied using molecular dynamics (MD) based on a semi-empirical potential. Magnesium is used as a paradigm hcp metal and interatomic interactions are modeled with the standard embedded atom method potential developed by Liu *et al*. [14]. Since the structure of the twin nucleus is still unknown, two hypothetical twin nuclei were considered. A volume containing a twinned subdomain was created by (1) mimicking the analytical method of Eshelby's inclusion problem [15] and (2) introducing a dipole of partial dislocations in an otherwise perfect crystal as proposed by Wang *et al.* [8].

It is found that in both cases, a specific interface, which places face to face a prismatic and a basal plane, plays an important role in the growth of the initial twin nucleus. In the following, this interface will be referred to as a *prismatic/basal interface*, or *PB interface* in short. As discussed below, PB interfaces appear in several atomic-scale structures published in the literature, both experimental and numerical, although, to the best of the authors' knowledge, the specific structure, stability and potential role of this interface in twin growth has not been discussed so far.

The orientation of the simulation cell is shown in Fig. 1. Horizontal planes are $(\bar{1}012)$ twinning planes with the x-axis along the $[10\bar{1}1]$ twinning direction. The z-axis, perpendicular to the plane of the figure, is parallel to the $[1\bar{2}10]$ direction. Stacking in this direction is binary, as illustrated by the two colors used to visualize the atoms. The $[1\bar{2}10]$ direction is contained in a second twinning plane, the $(10\bar{1}2)$ plane, which is almost perpendicular to the horizontal $(\bar{1}012)$ plane (the angle between the two $\{\bar{1}102\}$ planes is $2\tan^{-1}(\sqrt{3}a/c) = 93.7°$, with $c/a$=1.623 for Mg). Moreover, the $[1\bar{2}10]$ axis also belongs to the basal plane and to a prismatic plane, both shown in Fig. 1. Periodic boundary conditions are used in all three directions in a cell of dimensions 76.1x38x0.32 nm$^3$ containing 40,000 atoms.

In the first case, a twinned region is created by a procedure inspired from Eshelby's method for solving the inclusion problem in an elastic medium [15]. A region bounded by $(\bar{1}012)$, prismatic and basal planes as illustrated in Fig. 1 is initially cut-out of the system. A mirror symmetry with respect to the $(\bar{1}012)$ twin plane is then applied to the atoms inside the cut-out region. The shape of the region after the symmetry operation is slightly different from the initial shape. In order to restore the initial shape and be able to put the atoms back in the simulation cell, a shear is applied parallel to the $(\bar{1}012)$ planes in the $[10\bar{1}1]$ direction. Energy minimization is then used to relax the configuration. If no strain is applied to the simulation cell, the twinned region is unstable, shrinks and disappears during the energy minimization, restoring the perfect hcp crystal. However, if the initial nucleus is sufficiently large and a shear strain $\gamma_{xy}$ is applied in the $[10\bar{1}1]$ direction parallel to the $(\bar{1}012)$ planes (through Lees-Ewards shifted periodic boundary conditions to maintain periodicity in the y-direction), the nucleus may be stabilized.

It is found that upon further increasing the applied shear strain, the nucleus becomes unstable and grows in the simulation cell. Several snapshots of the growth process are shown in Fig. 2. The initial nucleus, shown in Fig. 2(a), was stabilized by application of 5.5% simple shear (the shear stress in the cell is then ~1200 MPa). The top and bottom planes of the nucleus are $(\bar{1}012)$ twin planes, while laterally the nucleus is bordered by interfaces at about 45° from the horizontal twin planes. Inspection of the local atomic structure, as seen in the inset of Fig. 2(a), shows that both interfaces are equivalent and place face-to-face a prismatic and a basal plane, one in the matrix and the other in the nucleus. They are therefore PB interfaces as introduced above. It should be noted that the appearance of these interfaces does not depend on the initial shape of the cut-out region. For instance, a nucleus initially bordered by only primary and conjugate $\{\bar{1}102\}$ planes will also form PB interfaces upon relaxation.

To simulate the growth process, the nucleus was destabilized by applying an additional 0.5% shear strain such as to initiate transverse propagation of the twin domain. The growth process, shown in Fig. 2(b-e), was simulated using MD at 50 K, keeping the applied strain constant. This low temperature was chosen in order to focus on the growth process and avoid perturbations by other thermally-activated processes.

The twin grows in a conservative manner, both vertically (thickening in the ± y directions) and laterally (transverse propagation in the ± x directions). Transverse propagation of the twin is found to be faster than thickening, resulting in an overall extension of the twin in the $[10\bar{1}1]$ direction. We note that the nucleus initially grows in a direction at an angle with respect to the horizontal direction. There are therefore geometric steps on the $(\bar{1}012)$ interfaces to accommodate this slight deviation. In the lateral directions, the twin front is made of three interfaces: the conjugate $(10\bar{1}2)$ twinning plane and two PB interfaces, as indicated in Fig. 2(b). Since the rotation associated with the twin is close to 90°, but not exactly, the basal and prismatic planes across the PB interfaces should not be perfectly parallel. The two loci of intersection between the three planes bounding the twin front (i.e PB interface/conjugate $(10\bar{1}2)$ twin boundary and conjugate $(10\bar{1}2)$ twin boundary/PB interface) are therefore expected to be characterized by large incompatible elastic distortions due to the rapid transition in misorientations. These are likely to have a discrete representation equivalent to a partial disclination dipole. Such will be the subject to future studies.

Lateral motion of the twin front occurs via the nucleation and glide of steps, i.e. disconnections, along the PB interfaces. Instances of steps are visible in Fig. 2(b) and (c). Their height is $2c$, i.e. they comprise 2 basal planes. PB interfaces can be interpreted as slanted walls of twinning dislocations [8,16], aligned along either a basal or a prismatic plane. In this context, we see that PB interfaces do not move by the simultaneous motion of all the twinning dislocations it contains, but rather by the motion of some dislocations in the interface forming a step, which then runs along the interface. Thickening of the twin also occurs by the nucleation and propagation of steps along the upper and lower $(\bar{1}012)$ interfaces. Examples of steps are visible in Figs. 2(b) and (c).

Figs. 2(d-e) show the end of the growth process when both ends of the nucleus meet through the periodic boundary conditions. PB interfaces are again involved since the two twinned regions meet along the conjugate $(10\bar{1}2)$ twinning plane but are bordered by PB interfaces. After coalescence and retraction of the PB interfaces, a single twinned layer parallel to the $(\bar{1}012)$ plane is obtained. This layered structure is similar to that considered in the works of Serra and co-workers [9,10] on the motion of twinning planes perpendicular to their interfaces, i.e. their thickening. It is also similar to the structure observed experimentally in Ref. [17].

The second method used to generate a potential twin nucleus consists in introducing a dipole of particular partial dislocations within the system and letting the system relax via energy minimization. The Burgers vector of the partials is $\pm \boldsymbol{b_n}$ and the $\boldsymbol{b_n}$ vector is shown in Fig. 1. When the initial width of the dipole is larger than about 5 nm, the relaxed dipole structure consists of small twinned regions near the partial dislocation cores connected by a stacking fault in the horizontal $(\bar{1}012)$ plane, as shown in the inset of Fig. 3(a). When a $\gamma_{xy}$ shear strain on the order of a few percents is applied, the twinned regions coalesce, remove the stacking fault and generate the structure presented in Fig. 3(a). This nucleus has a similar shape as the previous nucleus, being bordered by $(\bar{1}012)$ planes and PB interfaces. The difference is that there are dislocation cores at both extremities of the nucleus. This nucleus is stable if the applied strain is reduced back to zero and becomes unstable if the applied strain is instead increased as in previous case. The growth process is shown in Figs. 3(b-d) using MD at 50K. Motion of PB interfaces is again involved, in particular near the partial dislocation cores. We note that in all our simulations, the partial dislocations remained sessile. As a result, the twin front has to bypass them and leaves behind stacking faults in prismatic planes of the twinned region, visible in Figs. 3(c) and (d). The latter could be an artifact of Liu's potential, which predicts a stable stacking fault in prismatic planes in contrast with first-principles calculations [18].

When the initial nucleus, whether the homogeneously twinned region (i.e. method 1) or the partial dislocation dipole (i.e. method 2), is too narrow, i.e. not extended enough in the x-direction, the nucleus grows mostly vertically and not in the x-direction, resulting in the formation of a twinned region along the conjugate $(10\bar{1}2)$ planes, that is almost perpendicular to the direction of the applied shear, a very counterintuitive result. It is therefore very important to consider wide enough initial nuclei.

If the misorientation across the PB interface is exactly 90°, a prismatic and a basal plane are perfectly parallel to the interface. The corresponding geometry is represented in Fig. 4(a). The elementary prismatic and basal unit cells placed against each other are rectangular with no lattice mismatch along their common $[1\bar{2}10]$ direction and a mismatch of $2(\sqrt{3}a - c)/(\sqrt{3}a + c) \approx 6.5\%$ in the orthogonal direction. Misfit dislocations are required to accommodate this mismatch. Since coherency is almost restored when 14 basal unit cells are placed against 15 prismatic unit cells ($14\sqrt{3}a/15c = 0.996$), we expect to have one misfit dislocation along the $[1\bar{2}10]$ direction every 14 basal unit cells. The Burgers vector of this dislocation will be contained in the interface, corresponding to a $c$ vector of the prismatic unit cell, in contrast with the twinning dislocations which Burgers vector intersects the interface at an angle close to 45°. We used a bicrystal with 14 prismatic unit cells on top of 15 basal unit cells with periodic boundary conditions to compute the interface energy. The resulting atomic structure is shown in Fig. 4(b), where the misfit dislocation, which has a wide core extended in the interface, is highlighted. The surface energy is found to be ~170 mJ/m$^2$, which is above, but close to the twin energy along a $\{\bar{1}102\}$ plane, ~120 mJ/m$^2$. In addition, we computed the surface energy of a twinned bicrystal as a function of the angle $\theta$ between its interface and the $[10\bar{1}1]$ direction, as noted in the inset of Fig. 4(c). The energy curve is shown in Fig. 4(c) for $\theta$ between 0° and 93.7°, i.e. between the two conjugate $\{\bar{1}102\}$ planes. Three cusps are visible. The ones at $\theta = 0°$ and $\theta = 93.7°$ correspond to the primary and conjugate twinning planes, which are equivalent with the present setting. The cusp at $\theta = \tan^{-1}(\sqrt{3}a/c) = 46.9°$ corresponds to the PB interface, confirming the relative stability of this interface. Actually, the PB interface is so stable that in the vicinity of the PB orientation, the interface decomposes into segments of PB interfaces connected by steps, as illustrated in the inset of Fig. 4(c). The stability of the PB interface can also be judged by comparison with the symmetric tilt grain boundary energies computed by Wang *et al.* [19], who found energies on the order of 250 mJ/m$^2$, with cusps for singular orientations down to between 100 and 150 mJ/m$^2$, which is only slightly below the PB interface energy.

Interestingly, PB interfaces appear in several atomic-scale structures published in the literature, but had never been mentioned or studied before. For instance, the simulated twin boundaries near ***b**$_n$* dipoles in Figs. 3 and 4 of Ref. [8] and Figs. 8 and 9 of Ref. [16] are composed of PB interfaces. Also, a PB interface is visible in the twinned region reported in Ref. [19], which nucleated at a symmetric tilt grain boundary after absorption of a four-dislocation pile-up. PB interfaces are also visible in the atomic-scale simulations of tractions along the c-axis performed by Qi et al. [20], which involve an interplay between twinning and dislocation plasticity. More importantly, Zhang *et al.* [21] reported very recently twin boundaries with large deviations from $\{\bar{1}102\}$ twinning planes and the inset of Fig. 2(b) of this work shows that the deviation is accommodated by a succession of PB interfaces in both of their equivalent perpendicular orientations. Such decomposition (faceting) is expected from the low energy of PB interfaces compared to other twin plane orientations.

In conclusion, we have seen that a particular interface, often seen but never reported, plays an important role in the growth of a twin. The twins considered here are hypothetical but the real nature of the twin nucleus is still unknown. We may expect that heterogeneous nucleation from grain boundaries is more likely [19], but this process has not been clearly explained nor simulated so far and in all cases, we can conclude from the present work that the front of a growing twin, independently of the initial nucleus, will be made of PB interfaces and $\{\bar{1}102\}$ twin planes because they are of much lower energy than all other twin interfaces. Their combined mobility will therefore control the overall kinetics of twin propagation.

The work was supported by the French National Research Agency under projects ATOPLAST and MAGTWIN.

Fig. 1 : Crystallography of the simulation cell. Horizontal planes are $(\bar{1}012)$ twinning planes with the x-axis along the $[10\bar{1}1]$ twinning direction. The z-axis perpendicular to the plane of the figure is the $[1\bar{2}10]$ direction, at the intersection between two twinning planes, the basal plane and a prismatic plane. Also noted in the top right corner of the figure are the partial Burgers vectors $b_n$ and $b_p$ forming a stable stacking fault in the horizontal $(\bar{1}012)$ plane and the twinning partial Burgers vector $b_{tw}$.

Fig. 2: Growth process of a twin from a defect-free twin nucleus simulated by MD at 50K. Colors scale with atomic energies, from blue at low energy to red at high energy (see text for details).

Fig. 3: Growth process of a twin from a dipole of partial dislocations, as proposed in Ref. [8]. The color code in the same as in Fig. 2. The dipole initial width is 20 nm. The dipole was initially relaxed under an applied shear strain of 5.4% and then subjected to an increment of 0.2%.

Fig. 4: Structure and energy of a prismatic/basal (PB) interface. (a) Basal and prismatic unit cells placed face-to-face across the PB interface, (b) Relaxed structure of the interface with the core of a misfit dislocation highlighted (same color code as in Figs. 2 and 3), (c) Energy of a twin interface as a function of its angle $\theta$ with respect to the horizontal $[10\bar{1}1]$ direction. An example of atomic configuration is shown in the inset. A cylindrical geometry was used with free boundary conditions on the outside. Energies were computed in an inner cylinder to avoid surface effects. Atoms not accounted for form a ring of light-blue atoms visible on the outside of the cylinder.

Fig. 1

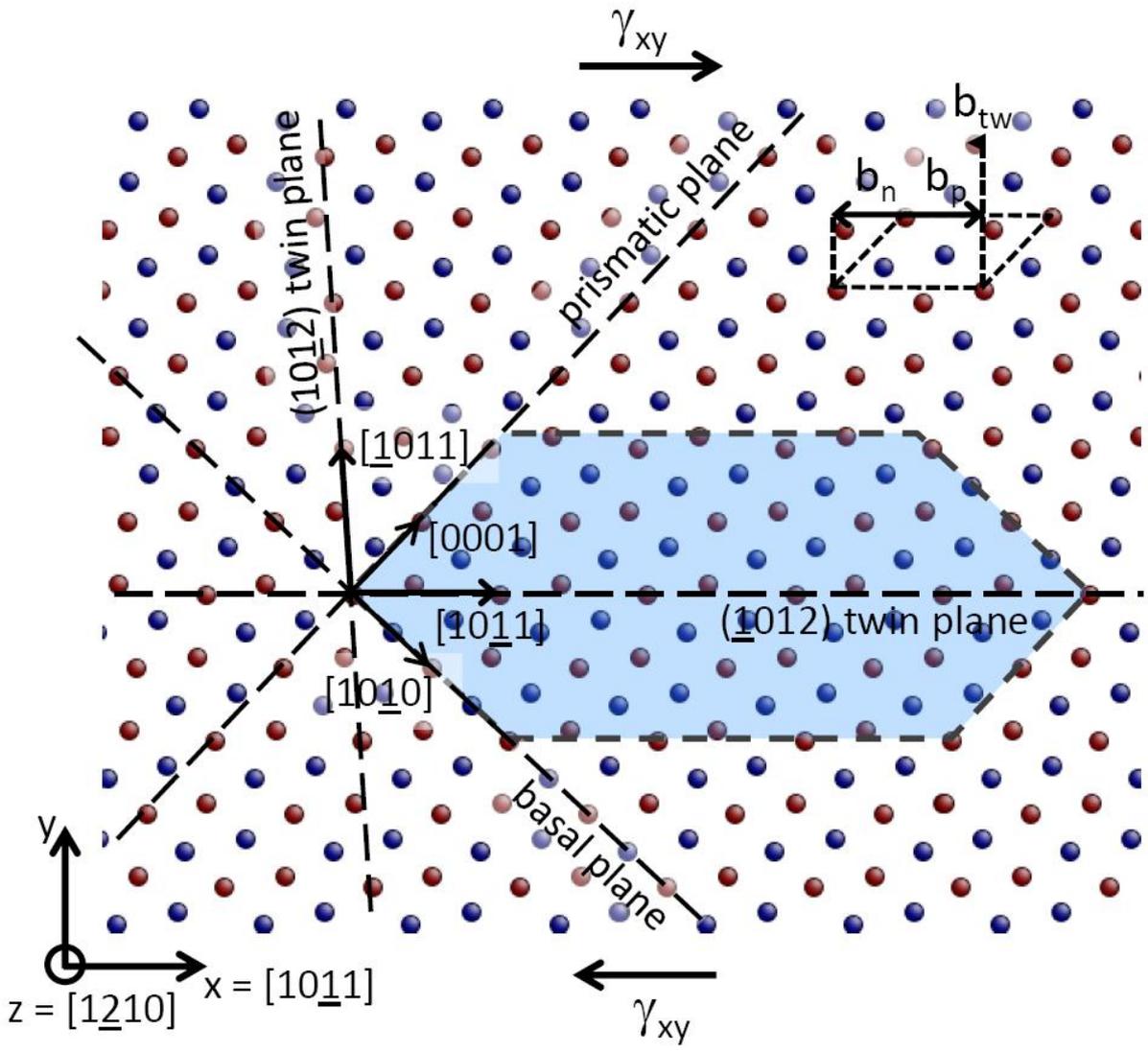

Fig. 2

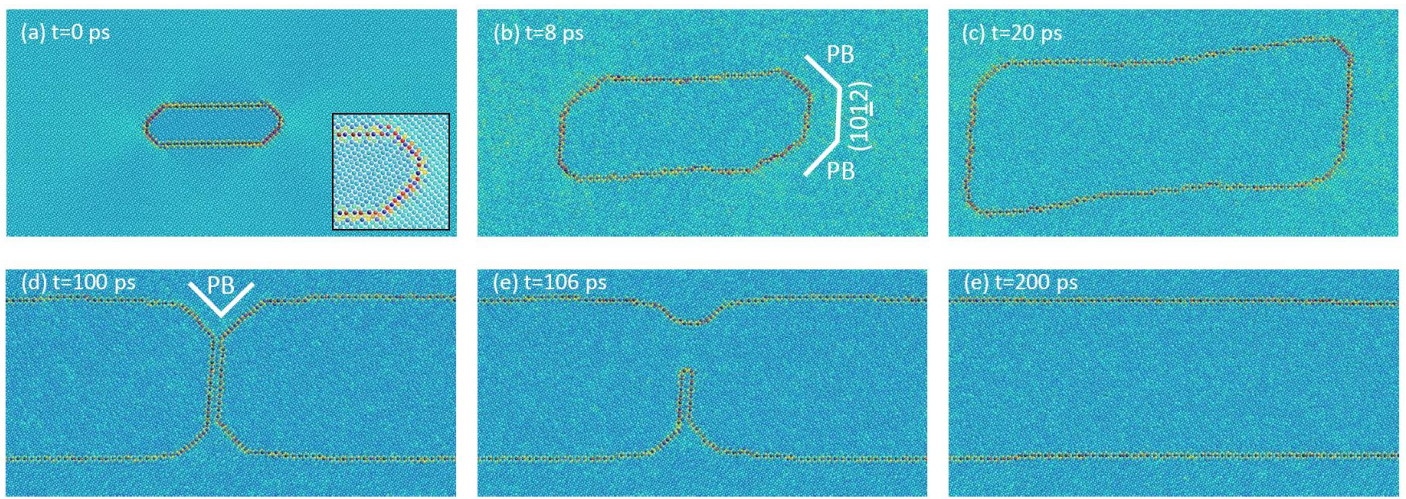

Fig. 3

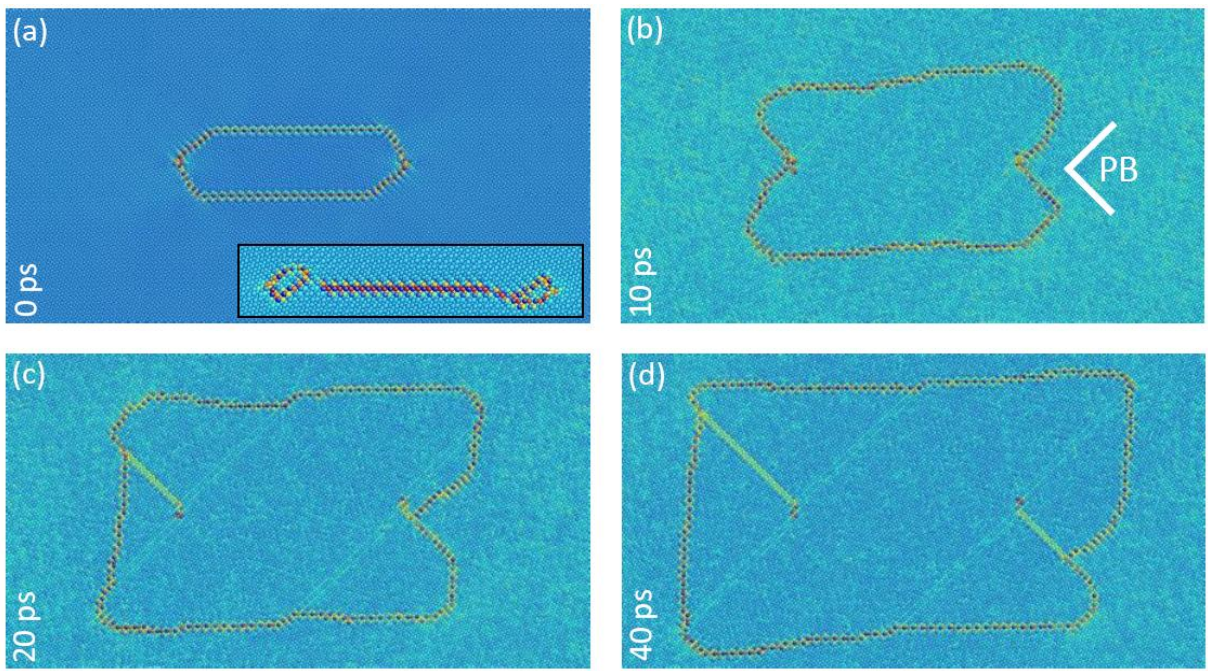

Fig. 4

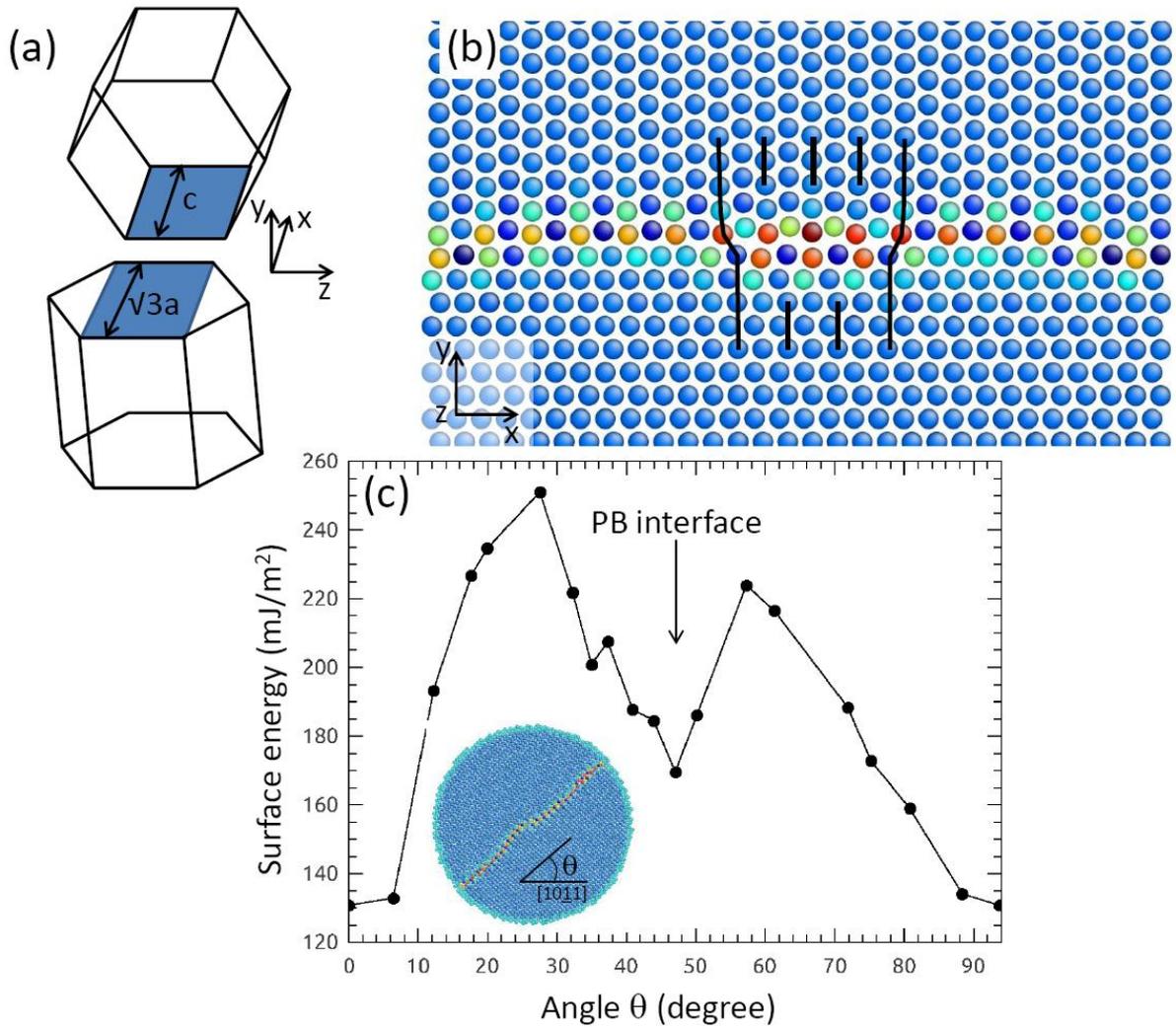